# A novel approach to modelling the properties of HEMTs operating in the saturation region

Kaiyuan Zhao[1,2,3], Guangfen Yao[1,2,3], Xiaoyu Cheng[1,2,3], Luqiao Yin[1,2,3], Kailin Ren[1,2,3,*] and Jianhua Zhang[1,2,3]

1 School of Microelectronics, Shanghai University, Shanghai 200444, China
2 Shanghai Collaborative Innovation Center of Intelligent Sensing Chip Technology, Shanghai University, Shanghai 200444, China
3 Shanghai Key Laboratory of Chips and Systems for Intelligent Connected Vehicle, Shanghai University, Shanghai 200444, China
* Correspondence: renkailin@shu.edu.cn

*Abstract*—Currently, the ASM-HEMT model, QPZD model and EPFL model are all based on the three-terminal potential as the core, and relate the electrical characteristics such as *I-V* and *C-V* to $V_d$, $V_s$ and $V_g$, so as to accurately build the HEMT model with high accuracy and fast convergence. However, there has not yet been a model based on three-terminal potentials that can quickly model the velocity saturation effect as well as the carrier concentration distribution and the electric field distribution inside the HEMT, which makes the existing models have to rely on a number of empirical parameters in the modelling process, which lacks the actual physical significance. In previous publications, models for the electric field, carrier concentration distribution based on the effective length of the gate were presented. In this paper, the model proposed in previous publications is improved to enable: (1) the calculation of the current $I_{ds}$ without relying on the Newton iterative method with fast simulation convergence behavior; (2) The $V_{dsat}$ when the HEMT reaches saturation at different $V_{go}$ is redefined instead of using $V_{dsat} = V_{go}$; (3) The expression of the *v-E* relationship is redefined relying on the different transport of carriers, which solves the problem of the large model error of the electric field distribution in the region below the gate, and makes the model's accuracy of the *I-V* characteristic further improved. The model was validated by characterising the *I-V* and *E-V* of the HEMT through TCAD simulation with RMSE below 5%.

*Key words*-ASM-HEMT model, QPZD model, v-E relationship, Vdsat calculation, electric field distribution.

## I. INTRODUCTION

Physically-based compact models capture the real transport of carriers inside the HEMT by combining the device size parameters to establish the physical characteristics of the HEMT such as *I-V* and *C-V*. Relying on the advantages of high accuracy, strong convergence, and good size extensibility, the physically-based model builds a strong bridge between the device process and circuit design, and is favoured by researchers. Currently, there are four dominant physically based HEMT models: ASM-HEMT [1][2], MVSG [3], EPFL [4] and QPZD [5][6] model. By accurately capturing the various physical effects inside the HEMT, all the above compact models can excellently model the electrical properties of the HEMT from different dimensions and application scenarios.

Modelling the electrical characteristics of HEMTs in the constant current region is necessary for requirements in analog circuit design. However, current models are still unable to model the electrical characteristics of the saturation region in a physical manner. The ASM-HEMT model uses ($V_{go}$-$V_{dsat}$) instead of $V_{go}$ to calculate the carrier concentration on the drain side when a large drain voltage ($V_d$) is applied [7]. However, $V_{dsat}$ is calculated by relying entirely on empirical expressions and cannot realistically characterise the pinch-off effect that occurs when $V_d$ is applied. Also, the physical meaning of $V_{dsat}$ is not the corresponding $V_d$ when the device reaches saturation. The QPZD model currently relies only on the channel length modulation factor to calculate the current corresponding to a HEMT operating in the saturation region, again lacking practical physical meaning [8]. Although these ways of modelling the saturated zone are fast and concise, they limit the accuracy and dimensional extensibility of the model.

In a previous publication [9], we presented a distribution model for the internal physical properties of T-gate HEMT based on ASM-HEMT and QPZD model. The model is centred on the gate effective length ($L_{g,eff}$) when the HEMT is operating in the saturation region, and in addition to the basic *I-V* and *C-V* models, a complete distribution model for the transverse electric field $E_x$ and the carrier concentration $n_s$ is also developed. However, the model still has deficiencies. For example, the established *I-V* characteristic model relies on the Newton's iterative method of calculation, which is unable to obtain explicit expressions; the $V_{dsat}$ applied when the HEMT reaches the saturation state is calculated by $V_{dsat} = V_{go}$, which lacks of practical physical significance; and the model of the $E_x$ distribution when the HEMT operates in the saturation region has an obvious error under the gate, which will lead to the slight bias of the $V_d$.

In order to better model the electrical properties of HEMTs operating in the saturated state, the model presented in [9] is further improved. In this paper, a novel way to define the expression of the *v-E* relation for the carriers is proposed, which allows the calculation of $I_{ds}$ in the linear region without relying on the Newtonian iterative method, and an explicit expression for the $V_x$ distribution model is obtained. At the same time, an explicit physical definition and calculation method for $V_{dsat}$ is presented. This work gives a new way to model HEMT working in the saturation region without relying on empirical parameters by explicitly defining and calculating $V_{dsat}$ and $L_{g,eff}$.

## II. DEVICE STRUCTURE AND MODEL FOUNDATION

Fig. 1 illustrates a schematic of HEMT used in this model, and structural parameters as well as the material parameters used in the modeling are shown in Table I. Fig. 2 demonstrates the division of different regions of the HEMT. When the HEMT operates in the linear region, the carrier concentration under the gate gradually decreases along the channel direction as $V_d$ keeps increasing. At this point, the HEMT can be divided into three regions, the Source Neutral Zone (SNZ), the Drain Neutral Zone (DNZ), and the Intrinsic FET Zone (IFZ). The potentials on both sides of the IFZ region are defined as $V_{si}$ and $V_{di}$ [10][11]. When the HEMT is operated in the saturation region, the pinch-off phenomenon starts to appear, and the pinch-off region is called the space charge limiting zone (SLZ). The region where the depletion of carriers fills up the SLZ region is called the Charge Deficit Zone (CDZ). The potential at the junction of the SLZ and IFZ is defined as $V_p$.

| | | |
|---|---|---|
| S | G | D |
| Passivation layer | | Passivation layer |
| 20 nm AlGaN barrier | | |
| 500 nm i-GaN channel | | |
| 1.3 μm GaN buffer | | |
| SiC substrate | | |

Table. 1 HEMT structural parameters

| Symbol | Definition | Value |
|---|---|---|
| $L_g$, $L_{gs}$, $L_{gd}$ | Gate/gate to source/ gate to drain | 1 μm/2 μm/ 7 μm |
| $W$ | Device width | 50 μm |
| $\varepsilon_1$ | AlGaN dielectric constant | $8.8 \times 10^{-13}$ F/cm |
| $\varepsilon_2$ | GaN dielectric constant | $8.7 \times 10^{-13}$ F/cm |
| $\mu_0$ | Low-field mobility | 2170 cm$^2$/V·s |
| $E_c$ | Saturation electric field strength | $1 \times 10^4$ V/cm |
| $N_s$ | Polarisation charge density | $1.02 \times 10^{13}$ cm$^{-2}$ |
| $C_g$ | Gate capacitance | $4.4 \times 10^{-7}$ F/cm$^2$ |

Fig. 1. Schematic cross-section of the HEMT.

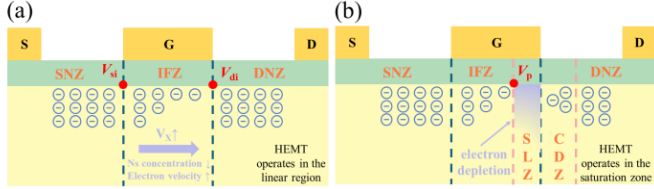

Fig. 2. Schematic diagram of the structure division when HEMT is operating in the (a) linear region, (b) saturation region.

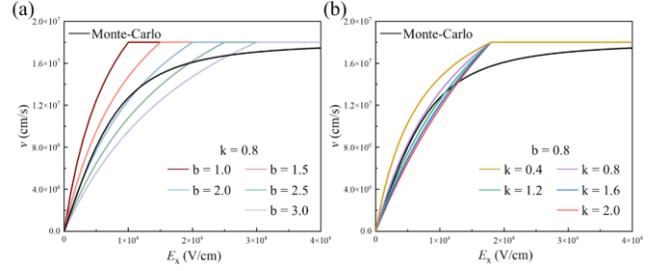

Fig. 3. Comparison of the curves after varying the (a) value of b, (b) value of k in the *v-E* expression with the Monte-Carlo curves.

$$E_f = V_{go} - \frac{2V_{th}\ln[1+\exp(V_{go}/(2V_{th}))]}{1/H(V_{go,eff}) + [C_G/(qD)]\exp(-V_{go}/(2V_{th}))} \quad (1)$$

$$n_S = \frac{C_g}{q}\left(V_{go} - \frac{E_f}{q} - V_X\right) \quad (2)$$

$$v(E) = \frac{\mu_0 E}{\sqrt{1+\left(E/E_c\right)^2}} \quad (3)$$

In previous publications, the ASM-HEMT model established in (1) was applied to determine the surface potential $E_f$ of GaN for different gate voltages [12]. Meanwhile, through the discussion of Eqs. (2)(3) for different operating states of HEMT, a compact model based on the effective length of the gate was established, which can be applied to model planar HEMT and T-gate HEMT. The shortcomings of this model will be improved in Section III.

### III. MODEL DESCRIPTION

#### A. Expression for the v-E relationship

In the established modelling framework, it is very difficult to physically model the HEMT when it is operating in the saturation region. This is because although the equation in (3) for the relationship between velocity and electric field holds under all conditions, as $V_d$ increases to bring the HEMT into the saturation region, carrier pinch-off occurs, and peaks in the transverse electric field ($E_x$) appear in the pinched region, rendering the Gradual channel approximation (GCA) no longer valid, and at the same time rendering (2) no longer applicable. The presence or absence of a pinch-off region is the only indication of whether the HEMT has entered the saturation region.

Previous publications [9] have defined the range in which eq. (2) applies when the HEMT operates in the saturation region. When the $E_x$ in the channel is less than $E_c$ in the region, both (2) and (3) apply. When the $E_x$ inside the channel is larger than $E_c$, both $n_s$ and $v$ are equated to constants and satisfy $I_{ds}=q \cdot W \cdot n_s \cdot v$. At this point, $v$ in the pinched region of the channel is everywhere defined as the maximum saturation velocity $v_{sat}$ the carriers can achieve. Although this approach can physically model a HEMT operating in the saturation region, there are still deficiencies in: (1) When the appropriate $u$ and $v_{sat}$ are chosen for TCAD simulation in the near-saturation state of $V_{go} = V_d$, the transverse electric field $E_x$ on the gate side against the drain is larger than $E_c$, indicating that the HEMT needs to be in a larger electric field strength before the pinch-off effect occurs. The original equivalent approach will cause errors in the calculation of $E_x$ below the gate and the calculation of $l_{g,eff}$; (2) the original equivalent approach makes the velocity of the carriers discontinuous at the point P, which causes errors in the modelling of the *I-V* characteristics; (3) the Monte-Carlo's *v-E* relation has a squared term in it, which creates difficulties in the subsequent derivation of the *I-V* characteristics as well as the explicit expressions for the *E-V* characteristics.

$$v(E) = \frac{\frac{k+1}{b} \cdot \mu \cdot E}{k + \frac{E}{b \cdot E_c}} \quad (4)$$

To address these shortcomings, a new expression for the *v-E* relationship is proposed in this paper, as shown in (4). It is worth noting that two dimensionless factors, k and b, are added into the original expression for the *v-E* relationship. In eq. (4), k determines the degree of inclination of the curve, while b determines at what electric field strength the carriers fully reach velocity saturation ($E = b \cdot E_c$, $v = v_{sat}$). Eq. (4) clearly defines where the carriers reach velocity saturation and there are no discontinuities in carrier velocity. At the same time, the absence of the v-E relation with squared terms brings convenience for subsequent calculations. In order to verify the accuracy of eq. (4), Fig. 3 shows the *v-E* relation curves after changing the values of k and b compared with the conventional Monte-Carlo curves [13] in eq. (3). The results show that by adjusting the values of k and b, eq. (4) can have a high agreement with the Monte-Carlo curve.

#### B. Calculation of $l_{g,eff}$ and $I_{ds}$

In previous publications, (3) was applied to the modelling of the *I-V* and *C-V* properties of the HEMT, as well as the $E_x$ and

$n_s$ distribution models. In this paper, the *v-E* relation satisfied by carriers is corrected from (3) to (4), and the new derivation is shown below.

Substituting eq. (4) into $I_{ds} = q \cdot W \cdot n_s \cdot v$ gives

$$I_{ds} = q \cdot W \cdot n_s \cdot v(E) = q \cdot W \cdot n_s \cdot \frac{\frac{k+1}{b} \cdot \mu \cdot E}{k + \frac{E}{b \cdot E_c}} \quad (5)$$

The $n_s$ in the access region is constant and numerically constant equal to the $n_s$ corresponding to $V_g = 0$. At this point, $E_x$ in the access region satisfies

$$E = \frac{\frac{b \cdot k}{k+1}}{\frac{q \cdot W \cdot n_s \cdot \mu}{I_{ds}} - \frac{1}{(k+1) \cdot E_c}} = \frac{b \cdot k \cdot E_c}{(k+1) \cdot \frac{I_{max}}{I_{ds}} - 1} \quad (6)$$

at this time, the potential $V_{si}$ on the gate side near the source and the potential $V_{di}$ on the gate side near the drain satisfy

$$V_{si} = \frac{b \cdot k \cdot E_c \cdot L_{sg}}{(k+1) \frac{I_{max}}{I_{ds}} - 1} \quad (7)$$

$$V_{di} = V_{ds} - \frac{b \cdot k \cdot E_c \cdot L_{dg}}{(k+1) \frac{I_{max}}{I_{ds}} - 1} \quad (8)$$

When the HEMT operates in the linear region, GCA applies everywhere below the gate. Substituting (2) and (4) into (5) yields

$$I_{ds} = q \cdot W \cdot \frac{C_g}{q} \cdot (V_{go} - V_x) \cdot \frac{\frac{k+1}{b} \cdot \mu \cdot E}{k + \frac{E}{b \cdot E_c}} \quad (9)$$

, simplifying eq. (9) gives

$$\left[\left(\frac{b \cdot k}{k+1}\right) \cdot \frac{1}{E} + \frac{1}{(k+1) \cdot E_c}\right] \cdot I_{ds} = \mu \cdot W \cdot C_g \cdot (V_{go} - V_x) \quad (10)$$

, substituting $E = dV/dx$ into (10) and simplifying (10) gives

$$\left(\frac{b \cdot k}{k+1}\right) dx = \left[\frac{\mu \cdot W \cdot C_g \cdot (V_{go} - V_x)}{I_{ds}} - \frac{1}{(k+1) \cdot E_c}\right] dV \quad (11)$$

, integrating both sides of the eq. (11) simultaneously, gives

$$\int_0^{L_g} \left(\frac{b \cdot k}{k+1}\right) dx = \int_{V_{si}}^{V_{di}} \left[\frac{\mu \cdot W \cdot C_g \cdot (V_{go} - V_x)}{I_{ds}} - \frac{1}{(k+1) \cdot E_c}\right] dV \quad (12)$$

At this point, solving (12) yields the expression satisfied by the $I_{ds}$ of the HEMT operating in the linear region, as follow

$$I_{ds} = \frac{\mu \cdot W \cdot C_g \cdot \left(V_{go} - \frac{V_{di} + V_{si}}{2}\right) \cdot (V_{di} - V_{si})}{\frac{b \cdot k}{k+1} L_g + \frac{V_{di} - V_{si}}{(k+1) \cdot E_c}} \quad (13)$$

When the HEMT operates in the saturation region, pinching effects begin to appear, and in the pinched region, the carrier velocity reaches $v_{sat}$, making GCA no longer applicable. However, in the IFZ region, (2) and (4) still apply. In order to equate the complex potential distribution in the IFZ, a new coefficient $a$ is introduced, and $a$ satisfies $V_{IFZ} = a \cdot b \cdot E_c \cdot l_{g,eff}$. At this point, the pinch-off point P satisfies the boundary conditions eq. (14), (15), and (16). Where, eq. (16) is obtained by substituting $v = v_{sat}$ and $n_s = C_g \cdot (V_{go} - V_p)/q$ for $I_{ds} = q \cdot W \cdot n_s \cdot v$.

$$E_P = b \cdot E_c \quad (14)$$

$$V_P = V_{si} + a \cdot b \cdot E_c \cdot L_{g,eff} \quad (15)$$

$$V_P = V_{go} - \frac{I_{ds}}{W \cdot C_g \cdot v_{sat}} \quad (16)$$

Substituting the boundary conditions (14) and (15) satisfied at point P into eq. (10), get

$$\left[\left(\frac{b \cdot k}{k+1}\right) \cdot \frac{1}{b \cdot E_c} + \frac{1}{(k+1) \cdot E_c}\right] \cdot I_{ds} = \mu \cdot W \cdot C_g \cdot (V_{go} - V_{si} - a \cdot b \cdot E_c \cdot L_{g,eff}) \quad (17)$$

, at this point, $L_{g,eff}$ can be written as

$$L_{g,eff} = \frac{V_{go} - V_{si} - \frac{I_{ds}}{E_c \cdot \mu \cdot W \cdot C_g}}{a \cdot b \cdot E_c} \quad (18)$$

, integrating the left and right sides of eq. (11), the upper limit of integration on the left side of the equation is changed from $L_g$ to $L_{g,eff}$, and the upper limit of integration on the right side of the equation is changed from $V_{di}$ to $V_p$, which gives

$$\int_0^{L_{g,eff}} \left(\frac{b \cdot k}{k+1}\right) dx = \int_{V_{si}}^{V_p} \left[\frac{\mu \cdot C_g \cdot (V_{go} - V_x)}{I_{ds}} - \frac{1}{(k+1) \cdot E_c}\right] dV \quad (19)$$

, solving eq. (19) yields the expression satisfied by the $I_{ds}$ of the HEMT operating in the saturation region, as follow

$$\frac{V_{go} - V_{si} - \frac{I_{ds}}{E_c \cdot \mu \cdot W \cdot C_g}}{a \cdot b \cdot E_c} = \frac{\frac{\mu \cdot W \cdot C_g}{I_{ds}} \cdot \left(V_{go} - \frac{V_c + V_{si}}{2}\right) \cdot (V_c - V_{si}) - \frac{V_c - V_{si}}{(k+1) \cdot E_c}}{\frac{b \cdot k}{k+1}} \quad (20)$$

In previous publications, it was proposed that the $V_d$ corresponding to different $I_{ds}$ can be obtained by varying the value of a and combining the electric field strengths in the SLZ and CDZ. It is worth noting that, compared to the $I_{ds}$ expression proposed in [9], in this paper, the correction of the *v-E* relation makes it unnecessary to rely on Newtonian iterative method for the solution of the HEMT operating in the linear region. Moreover, due to the continuous *v-E* relation, the $I_{ds}$ of the HEMT operating at the junction of the linear and saturation regions is made completely continuous, which greatly attenuates the current spikes occurring there.

*C. Presentation of $V_x$ expressions*

After solving for the $I_{ds}$ corresponding to the HEMT in the linear and saturation regions in the previous section, at this point, the corrected *v-E* relation can be further solved for the expression of the complex $V_x$ distribution in the IFZ. Simplification of eq. (10) yields

$$\frac{1}{E_x} = \frac{\mu \cdot W \cdot C_g \cdot (V_{go}-V_x)}{\frac{b \cdot k}{k+1} \cdot I_{ds}} - \frac{1}{b \cdot k \cdot E_c} \quad (21)$$

, and the solution of this differential equation is

$$x + M = \left[\frac{\mu \cdot W \cdot C_g \cdot V_{go}}{\frac{b \cdot k}{k+1} \cdot I_{ds}} - \frac{1}{b \cdot k \cdot E_c}\right] \cdot V_x - \frac{\mu \cdot W \cdot C_g}{2 \cdot \frac{b \cdot k}{k+1} \cdot I_{ds}} \cdot V_x^2 \quad (22)$$

The expression satisfied by $V_x$ is obtained by substituting the boundary condition $V_x = V_{si}$ at x = 0 into (22), as follow

$$x = \left[\frac{\mu \cdot W \cdot C_g \cdot V_{go}}{\frac{b \cdot k}{k+1} \cdot I_{ds}} - \frac{1}{b \cdot k \cdot E_c}\right] \cdot V_x - \frac{\mu \cdot W \cdot C_g}{2 \cdot \frac{b \cdot k}{k+1} \cdot I_{ds}} \cdot V_x^2 \\ - \left[\frac{\mu \cdot W \cdot C_g \cdot V_{go}}{\frac{b \cdot k}{k+1} \cdot I_{ds}} - \frac{1}{b \cdot k \cdot E_c}\right] \cdot V_{si} + \frac{\mu \cdot W \cdot C_g}{2 \cdot \frac{b \cdot k}{k+1} \cdot I_{ds}} \cdot V_{si}^2 \quad (23)$$

*D. Physical meaning and calculation of $V_{dsat}$*

Based on the derivation in Section B, the model proposed in this paper explicitly defines the point where the HEMT operates at the junction of the linear and saturation regions (whether or not a pinch-off effect occurs in the channel). After the pinch-off effect occurs, the electric field strength in the pinch-off region increases rapidly and the $I_{ds}$ saturates. The $V_d$ that corresponds to the point where the pinch-off effect happens to occur in the HEMT is defined as the $V_{dsat}$. It is worth noting that the $V_{dsat}$ proposed in this paper is different from that proposed in the ASM-HEMT model, and it has a clear physical meaning rather than relying on empirical fitting.

In eq. (20), which is satisfied when the HEMT is operating in the saturation region, the left and right sides of the equation are numerically equivalent to $L_{g,eff}$. Under the condition that $V_d = V_{dsat}$ and $L_g = L_{g,eff}$, eq. (20) can be written as

$$\frac{V_{go}-V_{si}-\frac{I_{ds}}{E_c \cdot \mu \cdot W \cdot C_g}}{a \cdot b \cdot E_c} = \frac{\frac{\mu \cdot W \cdot C_g}{I_{ds}} \cdot \left(V_{go}-\frac{V_c+V_{si}}{2}\right) \cdot (V_c-V_{si}) - \frac{V_c-V_{si}}{(k+1) \cdot E_c}}{\frac{b \cdot k}{k+1}} = L_g \quad (21)$$

In (21), the last two terms can be used to determine the $I_{ds}$ corresponding to $V = V_{dsat}$. At this point, substituting the value of $I_{ds}$ back into the first term determines the $a$ corresponding to $V = V_{dsat}$, and $V_{dsat}$ is given by

$$V_{dsat} = a \cdot b \cdot E_c \cdot L_g + V_{si} + V_{di} \quad (22)$$

## IV. MODEL VALIDATION AND DISCUSSION

TCAD simulations to obtain *I-V* characteristics and *E-V* characteristics are used to validate the model proposed in this paper. It is worth noting that the simulations in this paper are set up using the default values of the GaN material parameters. Fig. 4(a) plots the $I_D$-$V_D$ curves for the simulated and modelled values for $V_{go}$ = 1, 2, 3 and 4 V. The probability of a carrier being scattered is different at different $V_{go}$. The greater the carrier concentration, the stronger the scattering of the carriers, and the greater the electric field strength required for the carriers to reach velocity saturation. This also means that k and b need to be varied to adjust the *v-E* expression to simulate the

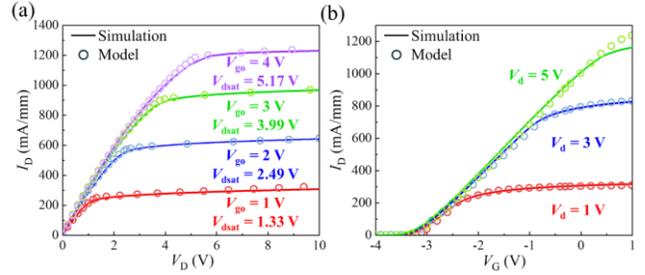

Fig. 4 (a) $I_D$-$V_D$ curve validation, (b) $I_D$-$V_G$ curve validation.

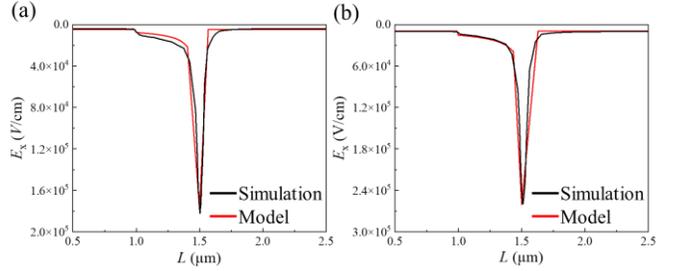

Fig. 5. Comparison of the $E_X$ distribution obtained by TCAD simulation and model, when planar HEMT is operating at (a) $V_{GO}$ = 2 V, $V_D$ = 4 V, (b) $V_{GO}$ = 4 V, $V_D$ = 8 V.

motion of carriers at different $V_{go}$. This effect leads to different $V_{dsat}$ at different $V_{go}$, and the larger the $V_{go}$, the larger the $V_{dsat}$, as shown in Fig. 6(a). Fig. 4(b) plots the $I_D$-$V_G$ curves for simulated and modelled values for $V_D$ = 1, 3, and 5V. The modelled values of $I_D$-$V_D$ and $I_D$-$V_G$ are well fitted to the simulated values with RMSE < 5%. Fig. 5 plots the TCAD simulation results and modelling results for $E_x$ at $V_{go}$ = 2 V, $V_d$ = 4 V and $V_{go}$ = 4 V, $V_d$ = 8 V. It is worth noting that the corrected *v-E* expression allows for more accurate modelling of the *E-V* curve by capturing the electric field (b·$E_c$) when the carriers reach saturation velocity. Compared to the modelling results in previous publication [9], the accuracy of the $E_x$ distribution model in the IFZ region is greatly improved when the HEMT is operating in the saturation region.

## V. CONCLUSION

In this paper, a new modelling approach for HEMTs operating in the saturation region is proposed. By rewriting the *v-E* expression, the $V_{dsat}$ when the HEMT reaches the saturation region and the $L_{g,eff}$ when it operates in the saturation region can be explicitly defined and calculated without relying on any empirical parameters. The model can accurately establish the *I-V* and *E-V* characteristic curves of the HEMT with RMSE <5%, as verified by simulation with TCAD.

Knowledgement
This work was supported by the Special Funds for Promoting High-quality Industrial Development in Shanghai under Grant JJ-ZDHYLY-01-23-0004.